\RequirePackage{fix-cm}
\documentclass[twocolumn]{svjour3}     

\smartqed  

\usepackage{multicol}

\usepackage{graphicx}
\usepackage{multicol,lipsum}
\usepackage{amsmath}
\usepackage{dcolumn} 
\usepackage{enumitem}
\usepackage{graphicx}
\usepackage{dcolumn}
\usepackage{array,multirow}
\usepackage{bm}
\usepackage{amsmath}
\usepackage{epstopdf}
\usepackage{amsfonts}
\usepackage{amssymb}
\usepackage{mathrsfs}
\usepackage{epsfig}
\usepackage{tabularx}
\usepackage{cite}
\usepackage[dvipsnames]{xcolor}
\RequirePackage[colorlinks,citecolor=blue,urlcolor=blue,linkcolor=blue]{hyperref}
\newcommand{\be}{\begin{equation}}
	\newcommand{\ee}{\end{equation}}
\newcommand{\beq}{\begin{eqnarray}}
	\newcommand{\eeq}{\end{eqnarray}}

	\usepackage{appendix}

\begin{document}

\title{Adiabatic analysis of the rotating BTZ black hole}

\author{Mohsen Fathi \and Samuel Lepe \and
        J.R. Villanueva 
}

\institute{
Mohsen Fathi \and J.R. Villanueva 
\at
    Instituto de F\'isica y Astronom\'ia,\\
    Universidad de Valpara\'iso, \\
    Avenida Gran Breta\~na 1111, Valpara\'iso, Chile
\and 
Samuel Lepe \at
              Instituto de F\'{\i}sica,\\ Pontificia Universidad de Cat\'{o}lica de
Valpara\'{\i}so,\\ Avenida Brasil 2950, Valpara\'{\i}so, Chile         
\and
Mohsen Fathi 
\at
    \email{\textcolor{Blue}{\href{mailto:}{mohsen.fathi@postgrado.uv.cl} }}
\and
Samuel Lepe 
\at
    \email{\textcolor{Blue}{\href{mailto:}{samuel.lepe@pucv.cl} }}
\and
J.R. Villuanueva
\at
    \email{\textcolor{Blue}{\href{mailto:}{jose.villanueva@uv.cl} }}
}

\date{Received: date / Accepted: date}
\maketitle

\begin{abstract}
In this paper we analyze some interesting features of the thermodynamics of the rotating BTZ black hole from the Carath\'{e}odory axiomatic postulate, for which, we exploit the appropriate Pfaffian form. The allowed adiabatic transformations are then obtained by solving the corresponding Cauchy problem, and are studied accordingly. Furthermore, we discuss the implications of our approach, regarding the the second and third laws of black hole thermodynamics. In particular, the merging of two extremal black holes is studied in detail.


\end{abstract}

\tableofcontents

\section{Introduction}
\label{intro}
The (2+1)-dimensional topological gravity has appeared to be of interest ever since the late twentieth century, especially because it can be used as a toy model for quantum gravity; the property that was discovered after the arguments presented regarding the possible connections between the (2+1)-dimensional gravity and the Chern-Simons theory \cite{Achucarro:1986,Witten:1988}. In the same realm, the Chilean physicists, Ba\~nados, Teitelboim, and Zanelli (BTZ), proposed a black hole solution in the $\rm{SO}(2,2)$ gauge group with a negative cosmological constant, that resembled, remarkably, the properties of the $(3+1)$-dimensional  Schwarzschild and Kerr black holes \cite{Banados:1992wn}, and hence, had a reliable physical significance. This solution was given further refinements, corrections and generalizations \cite{Banados:1992gq,Carlip:1995qv,Banados:1998gg,Cataldo:2000qi,AyonBeato:2004if}, and later, Witten  calculated the entropy of the BTZ black holes \cite{Witten:2007}. This solution, in fact, has gone through numerous investigations, form different perspectives. For instance, the geodesic structure of the uncharged BTZ black holes \cite{Cruz_1994}, the scattering process of test particles \cite{Gamboa:2000uc,Lepe:2003na}, the quasi-normal modes \cite{Cardoso:2001hn,Birmingham:2003wa,Crisostomo:2004hj,Setare:2003hm}, the hydrostatic equilibrium conditions for finite distributions \cite{Cruz:1994ar}, and solutions for fluid distributions matching the exterior BTZ spacetime \cite{Garcia:2004jz,Garcia_cuath,COV_2004BTZ,Gundlach:2020ovt}. Also, by considering a non-constant coupling parameter with the energy-scale, the scale-dependent version of the BTZ solution has been developed and discussed in Refs. \cite{Rincon:2018I,Rincon:2018II,Rincon:2019zxk,Fathi:2019jid,Rincon:2020izv}. It has been also argued that, if the energy-momentum complexes of Landau-Lifshitz and Weinberg are employed for a rotating BTZ black hole, the same energy distribution is obtained from both prescriptions \cite{Vagenas:2004zt}. This spacetime has been also generalized regarding the inclusion of terms related to the non-linear electrodynamics \cite{Cataldo_Garcia,Cataldo_Salgado} and the conformal group \cite{Carlip:2005zn}. Furthermore, regarding the black hole thermodynamics, the BTZ black holes have been investigated in terms of their critical behavior and phase transitions, by evaluating their equilibrium thermodynamic fluctuations in Ref.~\cite{Cai:1996df}, where it is shown that the extremal BTZ black hole with angular momentum serves as the critical point, and the density of states in the micro- and grand-canonical ensembles has been calculated in Ref.~\cite{Banados:1998ta}. For the case that the cosmological constant is considered as a thermodynamic parameter, the Kerr-(Anti-)de Sitter and the BTZ black holes have been compared Ref.~\cite{Wang:2006eb}. Furthermore, the quantum corrections to the enthalpy and the equation of state of the uncharged BTZ black holes were studied in Ref.~\cite{Dolan:2010ha}. In Ref.~\cite{Sarkar_2006}, a
general class of BTZ black holes is studied regarding the Ruppeiner geometry of the thermodynamic state space, and it is found that, this geometry is flat for both the rotating BTZ and the BTZ-Chern-Simons black holes, in the canonical ensemble. However, a non-zero scalar curvature is introduced to the thermodynamic geometry, when thermal fluctuations are included. In fact, this establishment of {\textit{geometrothermodynamics}}, as well that introduced in Ref. \cite{Quevedo:2008ry}, is a formalism that is used to designate a flat two-dimensional space of equilibrium states, which is endowed with a thermodynamic metric. This way, the space allows for the thermodynamic interaction, free of any kind of singularities (phase transitions). A more generalized consideration of geometrothermodynamics is available in Ref.~\cite{Akbar:2011qw}, where the thermodynamics of the charged BTZ black hole is investigated in the 
context of the Weinhold and Ruppeiner geometries. There, it is shown that these geometries cannot describe, completely, the black hole thermodynamics and the corresponding criteria for the electric charge. To solve this problem, in Ref. \cite{Hendi2}, a new metric (the HPEM metric) was introduced through a specific formalism, and this way, the corresponding Ricci scalar was shown to be able to bring together, different types of phase transitions. In Ref. \cite{Hendi1}, it is proved that the HPEM metric gives a consistent picture to study the thermodynamics of the BTZ black holes. The inclusion of quantum scalar fields in the study of black hole thermodynamics, was done significantly in Ref. \cite{Singh_2014}, for the case of static BTZ black hole, which lead to the introduction of entanglement thermodynamics for mass-less scalar fields. It has been also shown that the thermodynamics of BTZ black holes can be deformed in the context of gravity's rainbow, however, the Gibbs free energy remains unchanged \cite{Alsaleh}. The gravity's rainbow has been also exploited to study the black hole heat capacity and phase transition of BTZ black holes in Refs.~\cite{DEHGHANI2018351,Liang:2019jnj}. There are also some other extensions of the BTZ black holes that are derived in alternative theories of gravity. For example, in Ref. \cite{Camci:2020yre}, the Noether symmetries of the rotating BTZ black hole in $f(R)$ gravity has been used to generate new BTZ-type solutions. Along the same efforts, in Ref. \cite{Chougule:2018cny}, some thermodynamic aspects of the BTZ black holes, such as the Carnot heat engine, are studied in the context of massive gravity. Also, the authors in Ref. \cite{Bravo-Gaete:2014haa}, consider the Horndeski's action as the source field of the BTZ black hole, and reduce it to the common Einstein-Hilbert action including a cosmological constant. This way, they regain the usual three-dimensional Smarr formula by exploiting a scaling symmetry of this reduced action. Moreover, the rotating BTZ black holes, have been shown to exhibit no kind of superradiance, if the considered Dirac fields vanish at infinity \cite{Ortiz:2018ddt}.

There is also an interesting issue, regrading the $(2+1)$-dimensional {\it exotic} black hole solutions, whose spacetime metric has the same form as that of the BTZ, however, with mass and angular momentum being reversed in their roles ~\cite{Townsend:2013ela}. For this particular case, the entropy is proportional to the length of the inner horizon. The inner and outer horizons, in fact, limit the propagation of radial geodesics. Therefore, they can be thermally quantized only beyond the horizons. Accordingly, It has been shown in Ref. \cite{Kiselev:2005jb} that, the entropy of the BTZ black holes is in agreement with the Bekenstein-Hawking formula, and the particles retain their quantum ground level in the BTZ spacetime.

The reason of presenting such, relatively long, introduction is to highlight the interest of the scientific community in scrutinizing the thermodynamic of the $(2+1)$-dimensional, and in particular, the BTZ black holes. In this paper, based on the same interest, we study some thermodynamic aspects of the rotating BTZ black hole. Specifically, we base our discussion on the Carath\'{e}odory postulate of adiabatic inaccessibility \cite{caratheodory09}, that ensures the integrability of the Pfaffian form, $\delta Q_{{\rm rev}}$, that represents the infinitesimal heat exchanged reversibly. 
This approach allows for constructing a proper thermodynamic manifold by means of foliating the adiabatic surfaces that satisfy the Pfaffian equation $\delta Q_{{\rm rev}} = 0$. For the case of the (3 + 1)-dimensional black holes, this type of construction has been studied in detail in Refs. \cite{Belgiorno:2002BHCth,Belgiorno_Cacciatori,Belgiorno_Martellini} which gives rise to the isoareal transformations, i.e., the transformations between the black hole states with the same areas.
On the other hand, for the (2 + 1)-dimensional black holes, the adiabatic transformations correspond to the isoperimetral transformations between states that reside in the non-extremal manifold.

In order to elaborate on this, in Sect. \ref{sec2}, we introduce the rotating BTZ solution, and the way we approach based in the Carath\'{e}odory postulate. In Sect. \ref{AIT}, we study the allowed adiabatic transformation for the thermodynamic states, given the corresponding analytical solutions that constitute an adiabatic hypersurface. The second law of black hole thermodynamics is then seen in more details in Sect. \ref{sec4}, where we consider the scattering of two extremal BTZ black holes. We summarize the results in Sect. \ref{finrem}.


\section{The rotating BTZ black hole and its thermodynamics}\label{sec2}

{The (2+1)-dimensional, uncharged, black hole solution with a negative cosmological constant $\Lambda=-\ell^{-2}$, is obtained from the action
\begin{equation}
\label{actbtz}I=\frac{c}{2 \pi G} \int \sqrt{-g} \left[R+2\,\ell^{-2}\right] {\rm d}^2x\,{\rm d}t+\mathcal{B},
\end{equation}
where $\mathcal{B}$ is a surface term \cite{Banados:1992wn,Banados:1992gq}.
 For the stationary circular symmetry, the corresponding spacetime metric is given in terms of the coordinates $-\infty<t<\infty, 0<r<\infty$, and $0\leq\phi\leq 2\pi$, and can be written as
\begin{equation}\label{lineelans}
\mathrm{d}s^2 = -N^2(r) c^2 \mathrm{d}t^2 + N^{-2}(r) \mathrm{d}r^2 + r^2 \left[N^{\phi}(r) c\,{\rm d}t+\mathrm{d}\phi\right]^2,
\end{equation}
in which, the square lapse function and the angular shift are given, respectively, by
\begin{subequations}\label{lapshift}
\begin{align}
 & N^2(r)=-\frac{G M}{c^2}+\frac{r^2}{\ell^2}+\frac{G^2  J^2}{4 c^6 r^2},\\
 & N^{\phi}(r)=-\frac{G J}{2 c^3 r^2},
\end{align}
\end{subequations}
where $M$ and $J$, indicate the mass and the angular momentum of the black hole.}
{This spacetime possesses an inner ($r_-$), and an event ($r_+$) horizon, that are located at
\begin{equation}
	r_{\pm }=\frac{c\, \tau_{\pm}\left(\mathcal{M}, \mathcal{J}\right)}{\sqrt{2}},\label{eq1}
\end{equation}
where
\begin{equation}
\label{tt}\tau_{\pm}\left(\mathcal{M}, \mathcal{J}\right)=\sqrt{\mathcal{M} \pm\sqrt{\mathcal{M}^2-\mathcal{J}^2}},
\end{equation}
and
\begin{subequations}\label{mj}
\begin{align}
    & \mathcal{M}\equiv \frac{M}{m_p\,\Omega_{\mathrm{ext}}^2},\\
    & \mathcal{J}\equiv\frac{t_p\,J}{\hbar\,\Omega_{\mathrm{ext}}}.
\end{align}
\end{subequations}
Here the subscript $"p"$ is referred to the Planck quantities in (2+1) dimensions\footnote{In the (2+1)-dimensional gravity, the gravitational constant $G$ has the physical dimension of $\left[\rm{length}^2/(\rm{mass} \times \rm{time}^2)\right]$ (see Ref.~\cite{Cruzlepe04}). Therefore, the Planck mass, length and time become $m_p=c^2/G$, $l_p=G \hbar/c^3$, and  $t_p=l_p/c$.}, and  $\Omega _{\mathrm{ext}}=c/\ell$ is the angular velocity of the {\it extremal black hole}. Note that, the physical dimension of $\mathcal{M}$ and $\mathcal{J}$, is [$\rm{time}^2$], while the function $\tau_{\pm}$ has the dimension of $[\rm{time}]$. Also, in the extremal case the relation $\mathcal{M}=\mathcal{J}$ is satisfied. }

{The Bekenstein-Hawking entropy formula, if applied to the BTZ black hole, gives the entropy proportional to the event horizon's perimeter $P_{\mathrm{\mathrm{bh}}}=2\pi r_+$ instead of its area $A_{\mathrm{bh}}$, as it is expected from the dimensional ground. Therefore
\begin{equation}
	S=\frac{k_B}{4}\frac{P_{\mathrm{bh}}}{\ell_p}=\frac{k_B}{4}\left( \frac{2\pi r_{+}}{c\, t_p}\right) = a \tau_+,\label{eq2}
\end{equation}%
where $S$ is the entropy, $k_B$ is the Boltzmann constant, and  $a=(\pi/\sqrt{8})(k_B/t_p)\approx 1.1 (k_B/t_p)$. Defining $\mathcal{S}\equiv S/a=\tau_+$, and using Eqs. (\ref{tt}) and (\ref{eq2}),  we obtain a Christodoulou-type mass formula, which relates the total mass (energy) $\mathcal{M}$ to the entropy and the angular momentum, in the following form:
\begin{equation}
	\mathcal{M}(\mathcal{S}, \mathcal{J})=\frac{1}{2}\,\mathcal{S}^{2}+\frac{1}{2}\, \frac{\mathcal{J}^{2}}{\mathcal{S}^{2}}.  \label{eq3}
\end{equation}%
We base our our study on the framework of Carath\'{e}odory's approach to thermodynamics, 
that postulates the integrability of the Pfaffian form $\delta Q_{{\rm rev}}$, representing the infinitesimal heat exchanged reversibly \cite{caratheodory09,buchdahl49,buchdahl49I,buchdahl49II,buchdahl54,buchdahl55,Landsberg1964ADO,Marshall78,boyling68,boyling72,pogliani,Belgiorno:2002BHCth,Belgiorno2002Bridge,Belgiorno:2002QTBH,Belgiorno_Martellini,Belgiorno_2003aNTL,Belgiorno_2003bNTL,Belgiorno_Cacciatori}. 
In particular, we assume that the so-called metrical entropy $\mathcal{S}$ and absolute temperature $\mathcal{T}$, exist. Therefore, we can write
\begin{equation}
    \label{pfor} \delta Q_{\mathrm{\mathrm{rev}}}= \mathcal{T}\,{\rm d}\mathcal{S},
\end{equation}where $\mathcal{T}\geq 0$ is an integrating  factor which satisfies
\begin{equation}
    \label{tempdef} \frac{\partial \mathcal{S}}{\partial \mathcal{M}}\equiv \frac{1}{\mathcal{T}}>0,
\end{equation}
so that
\begin{equation}
    \label{temp}\mathcal{T}(\mathcal{M}, \mathcal{J}) =\frac{\left(\mathcal{M}+\sqrt{\mathcal{M}^2-\mathcal{J}^2}\right)^2-\mathcal{J}^2}{\left(\mathcal{M}+\sqrt{\mathcal{M}^2-\mathcal{J}^2}\right)^{3/2}}.
\end{equation}
In fact, if we choose the pair ($\mathcal{M}, \mathcal{J}$) as the extensive, independent variables in the equilibrium thermodynamics (i.e. homogeneous functions of degree one), then the homogeneity of the system is reflected in the integrability of the Pfaffian form  
\begin{equation}\label{pf2}
    \delta Q_{\mathrm{rev}}={\rm d}\mathcal{M}-\mathcal{W}\, {\rm d}\mathcal{J},
\end{equation}
where $\mathcal{W}$ is the angular  velocity of the black hole, given by 
\begin{equation}
    \label{angvelbtz} \mathcal{W}(\mathcal{M}, \mathcal{J}) =\frac{\mathcal{J}}{\mathcal{M}+\sqrt{\mathcal{M}^2-\mathcal{J}^2}}.
\end{equation}
Therefore, it is straightforward to show that, under the scaling transformation $(\mathcal{M}, \mathcal{J})\mapsto (\lambda \mathcal{M}, \lambda \mathcal{J})$, we get $\delta Q_{\mathrm{rev}}\mapsto \lambda \delta Q_{\mathrm{rev}}$, which means that the Pfaffian form is homogeneous of degree one. Consequently, we have an Euler vectorial field, or a Liouville operator, as the infinitesimal  generator of the homogeneous transformations
\begin{equation}
    \label{eulervec}
    D=\mathcal{M} \frac{\partial }{\partial \mathcal{M}}+\mathcal{J} \frac{\partial }{\partial \mathcal{J}},
\end{equation}
using which, we obtain
\begin{equation}
    \label{homentr} D \mathcal{S}=\frac{1}{2}\mathcal{S},
\end{equation}
meaning that, $\mathcal{S}$ is homogeneous of  degree $1/2$. Similarly, the temperature is also homogeneous of degree $1/2$. 
Furthermore, it is straightforward to check that the angular velocity is a homogeneous 
function of degree zero, or $D \mathcal{W} = 0$, and therefore, it is an intensive  variable.}

{It is, naturally, tempting to address a comparison with the natural (3+1)-dimensional counterpart (i.e. the Kerr-(Anti-)de Sitter black hole). In fact, there are some differences between these cases that should be analyzed carefully. Furthermore, we have found a mathematical equivalence of a remarkable theoretical potential. However, for the sake the scope of this study, for now, we strive on presenting some immediate results of the above discussed concepts, and leave the aforementioned mathematical comparison to a future work.}


\section{The adiabatic-isoperimetral transformations}
\label{AIT}	

{An important result of the above approach is that it allows for the generation of a non-extremal manifold foliation. In fact, the non-extremal thermodynamic space is foliated by those submanifolds of co-dimension one, which are solutions of the Pfaffian equation  $\delta Q_{{\rm rev}}=0$ \cite{Belgiorno:2002BHCth}. 
}

{As stated above, for the non-extremal manifold ($\mathcal{T}>0$), the Pfaffian form is given by Eq. (\ref{pf2}). Accordingly, performing the changes of variable $x=\mathcal{M}^2$ and $y=\mathcal{J}^2$, we get
\begin{equation}
	\delta Q_{{\rm rev}}=\frac{1}{2 \sqrt{x}}{\rm d}x- \frac{1}{2 (\sqrt{x}+\sqrt{x-y})}{\rm d}y,
	\label{at3}
	\end{equation}
that respects the condition $x\geq y$.} {Thus, for the isoperimetral transformation $\delta Q_{\mathrm{rev}}=0$, that connects, adiabatically, the initial state $\mathbf{i}\equiv (x_i, y_i)$ to the final state $\mathbf{f}\equiv (x_f, y_f)$, the adiabatic trajectories are solutions to the  Cauchy problem 
\begin{subequations}
    \begin{align}
 &    \frac{{\rm d}y}{{\rm d}x}=1+\sqrt{1-\frac{y}{x}}, \label{cach1}\\
 &   y(x_i)=y_i, \label{cach2}
    \end{align}
\end{subequations}
with $y_i < x_i$.} 
{It is then straightforward to show that, the solutions to this problem are  
\begin{subequations}
    \begin{align}
  & \label{sol1} y_{a}(x) =  2\sqrt{\zeta_a}\,\sqrt{x}-\zeta_a,\\ 
  &  \label{sol2} y_b(x) = 2\sqrt{\zeta_b}\,\sqrt{x}-\zeta_b,
    \end{align}
\end{subequations}
where the constants $\zeta_{a,b}\equiv \zeta_{a,b} (x_i, y_i)$ are given by
\begin{equation}
    \label{zetaval}\zeta_{a,b}=2x_i-y_i\pm 2\sqrt{x_i(x_i-y_i)},
\end{equation}
with $x_i> y_i$.
\begin{figure}[h!]
	\begin{center}
		\includegraphics[width=85mm]{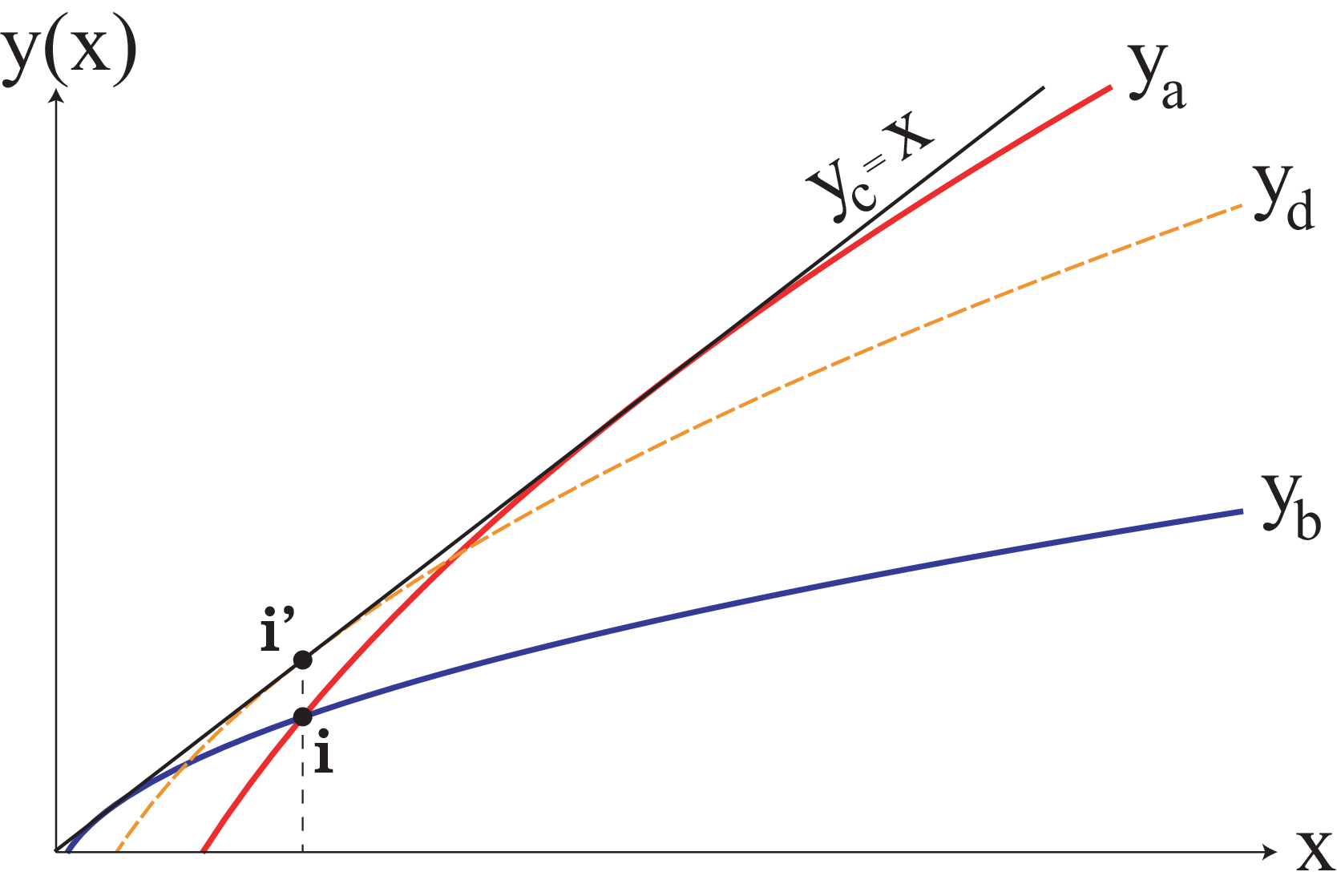}
	\end{center}
	\caption{The plots of the adiabatic solutions to the Cauchy problem, given in Eqs.~(\ref{sol1}), (\ref{sol2}), (\ref{sln1}) and (\ref{sln2}). The static black hole limit, for each case, is where the curves hit the $x$ coordinate, whereas, the extremal limit corresponds to the line $y=x$. 
	}
	\label{fig1}
\end{figure}}
{Each function vanishes at the point $(x_0, 0)$, with $x_0=\zeta_{a,b}/4$, which corresponds to the static BTZ black hole. The thermodynamic (extremal) limit, on the other hand, is reached at $(x_e, x_e)$, with $x_e=\zeta_{a,b}$ (see Fig. \ref{fig1}, showing both functions intersecting at the initial point $\mathbf{i}$). We will come back to these concepts later in this section.}

{Note that, it is important to be cautious about the conditions on the extremal submanifold, on which, the condition $\delta Q_{{\rm rev}}=0$ is still satisfied. This implies that the extremal submanifold is an integral submanifold of the Pfaffian form \cite{Belgiorno:2002BHCth,Belgiorno_Martellini}.} {In fact, considering the extremal point $\mathbf{i}'\equiv (x_i, x_i)$ as initial state, the Cauchy problem becomes
\begin{subequations}
    \begin{align}
  &  \label{cach3}\frac{{\rm d}y}{{\rm d}x}=1+\sqrt{1-\frac{y}{x}},\\
  & \label{cach4}y(x_i)=x_i,
    \end{align}
\end{subequations}
which allows for the two solutions
\begin{subequations}
    \begin{align}
    &  \label{sln1} y_c(x) = x,\\
    &  \label{sln2} y_d(x) = 2\sqrt{x_i}\, \sqrt{x}-x_i.
    \end{align}
\end{subequations}
Now that the solutions to the both cases of non-extremal and extremal cases have been given, it is of importance discussing their physical features, regrading the adiabatic processes.
In particular, the solution $y_a$ given by Eq. (\ref{sln1}), indicates that the extremal states are adiabatically connected to each other. However, the solution $y_b$ in Eq. (\ref{sln2}), presents a more complicated situation, because it connects, adiabatically, the non-extremal states with the extremal ones. This, in fact, poses a contradiction to the second law of thermodynamics, since it provides the possibility to construct a Carnot cycle with one hundred percent thermal efficiency, and this, violates the Ostwald's postulate of the second law. Furthermore, it would be possible to transform, completely, the heat into work, that is also in contrast with the second law.
To eliminate this singular behavior of the thermodynamic foliation, we assume that the surface $\mathcal {T} = 0$ is a leaf itself, that is, we exclude it from the set of solutions.} 
{Accordingly, by introducing a discontinuity in $\mathcal{S}$ between the extremal and non-extremal states, we construct a foliation of the thermodynamic variety, whose leaves are distinguished by
\begin{equation}\label{entrbtz}
  \mathcal{S}(\mathcal{M}, \mathcal{J}) = \left\{
             \begin{array}{ll}
             \mathcal{P}/4,& \quad \textrm{non-extremal states,} \\
             \medskip\\
             0,& \quad \textrm{extremal states.} \\
             \end{array} \right.
\end{equation}
The choice of the value $\mathcal{S}=0 $ for the extremal states, stems in some topological preferences \cite{Hawking:1994ii,Teitelboim:1994az}, and has been explicitly proposed by Carroll in Ref.~\cite{carroll_extremal_2009}. Nevertheless, it has been shown that this choice, is a particular case of a well-behaved area-dependent function, that can opt non-zero values \cite{lemos_entropy_2016}. In particular, the thin shells (rings) in the (2+1)-dimensional gravity, can change their entropy values during their evolution to a black hole (see Refs.~\cite{lemos_entropy_2014,lemos_thermodynamics_2015,lemos_unified_2017}).
}
\begin{figure}[h!]
	\begin{center}
		\includegraphics[width=85mm]{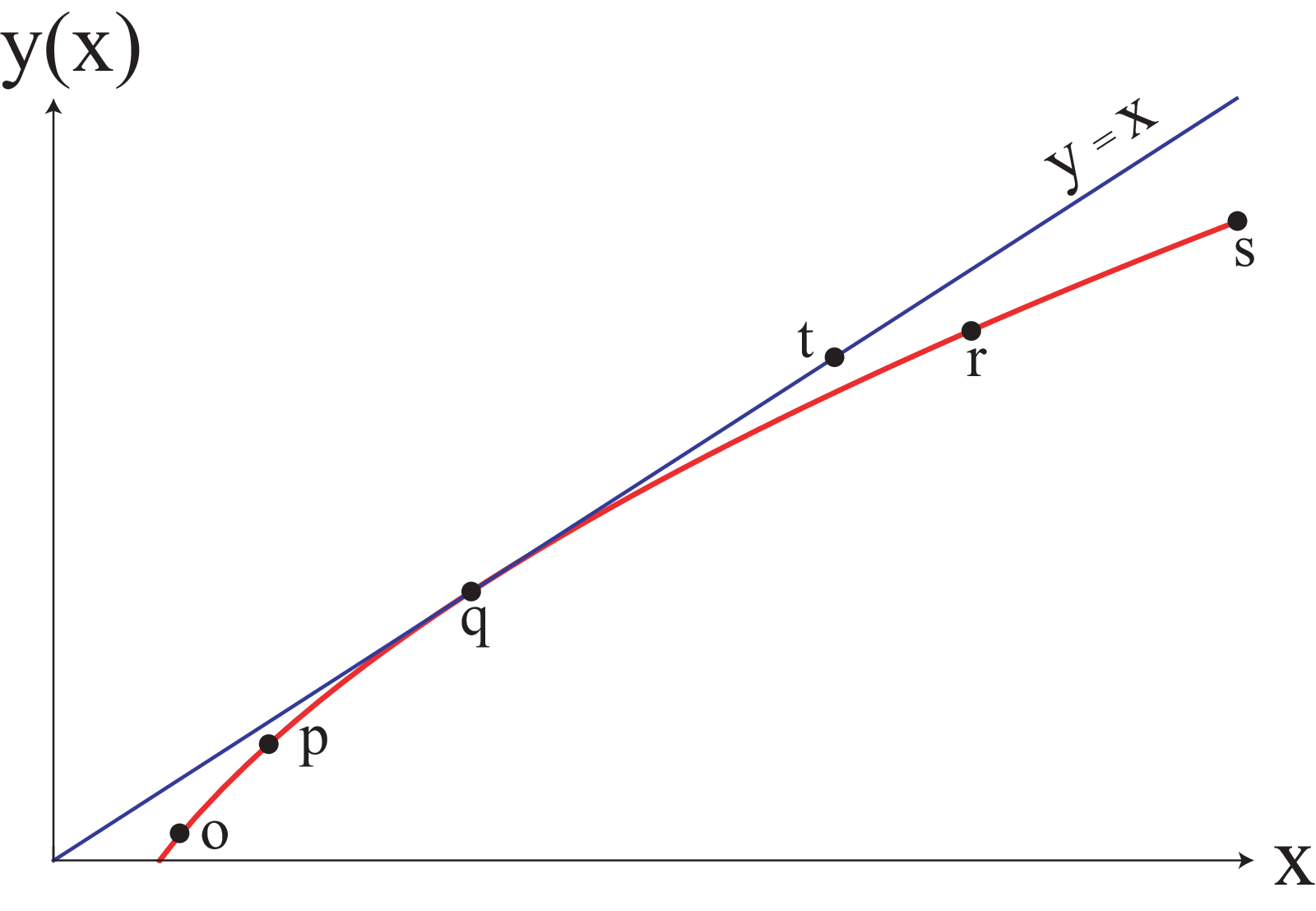}
	\end{center}
	\caption{The adiabatic solutions to the Cauchy problem and the extremal limit. In order to avoid violation of the second law, the only allowed processes are o $\leftrightarrow$ p; r $\leftrightarrow$ s; q $\leftrightarrow$ t. The processes p $\leftrightarrow$ q; q $\leftrightarrow$ r are, on the other hand, prohibited.} 
	\label{fig2}
\end{figure}

\begin{figure}[h!]
	\begin{center}
		\includegraphics[width=85mm]{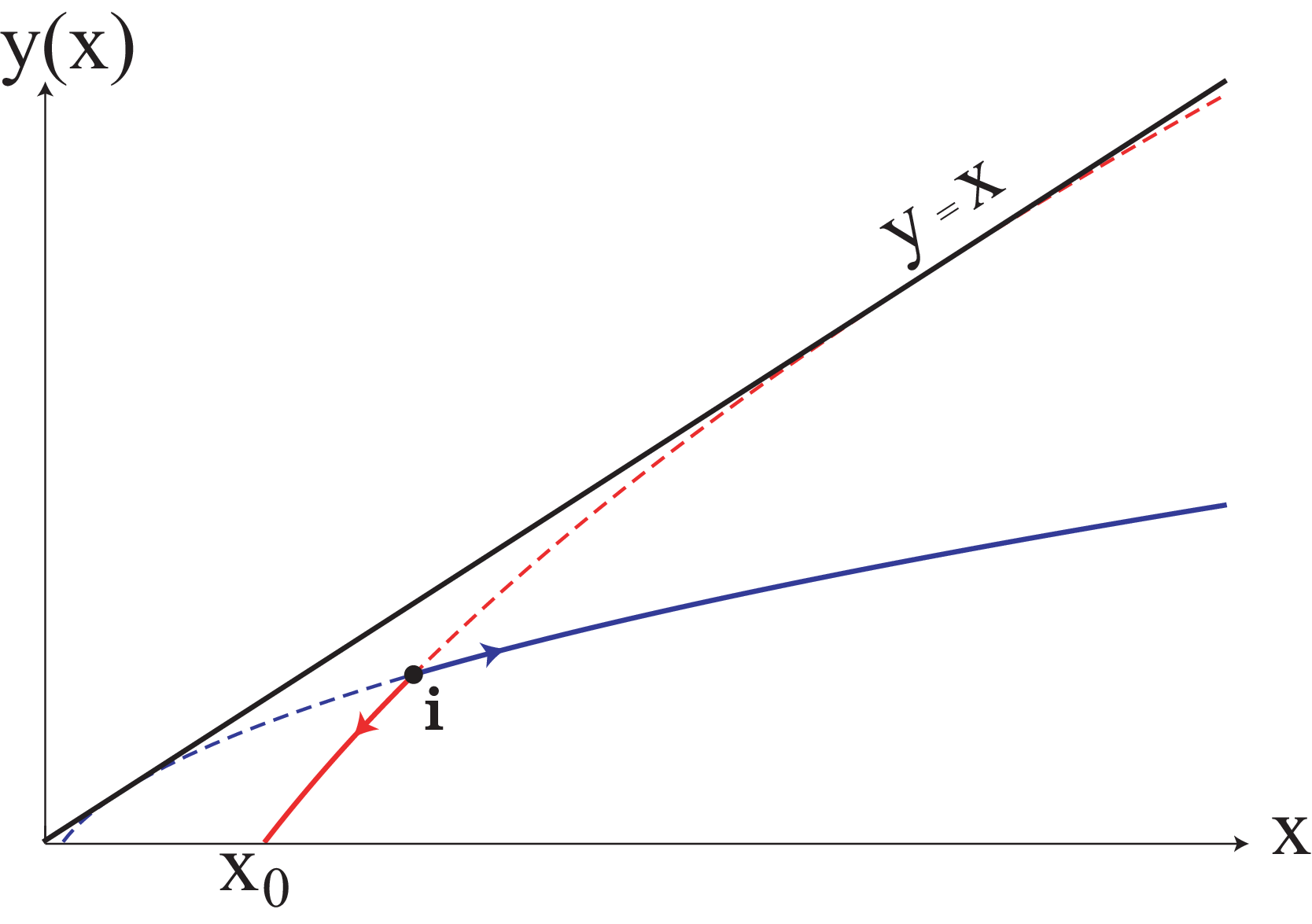}
	\end{center}
	\caption{The physically allowed solutions to the Cauchy problem for the BTZ black hole. The initial state $\mathbf{i}\equiv (x_i, y_i)$, with $y_i<x_i$, can be connected, adiabatically, to the final state $\mathbf{f}\equiv (x_f, y_f)$, with $y_f<x_f$, following the path $y_a$ (red curve), if $x_f<x_i$, and the path $y_b$ (blue curve), if $x_f>x_i$.
	}
	\label{fig3}
\end{figure}

{We can now establish a criteria for the  physically acceptable solutions, based on the results obtained above:
\begin{enumerate}
    \item Due to the homogeneity of the extensive variables ($\mathcal{M}, \mathcal{J})$ or ($x, y$), every adiabatic process must satisfy
    \begin{equation}
        \label{c1} \frac{{\rm d}y}{{\rm d}x}>0.
    \end{equation} 
    \item An initial state belonging to the submanifold $\mathcal{T}>0$, can only be adiabatically connected to another state, if it neither belongs to the submanifold $\mathcal{T}=0$, nor it passes through it.
    \item In the neighborhood of any equilibrium state of the system, there exists states that are inaccessible by the reversible adiabatic processes (Carath\'{e}odory postulate) \cite{caratheodory09,buchdahl49,buchdahl49I}.
\end{enumerate}
Condition 1, is nothing but the result of expressing the thermodynamic system in terms of the extensive variables, which are, of course, homogeneous of degree one. 
Condition 2, ensures satisfaction of the second and third laws of thermodynamics. From the geometric point of view, this guarantees that the black hole topology does not change. The above statements have been visualized, qualitatively, in Fig \ref{fig2}. There, we have exemplified the allowed processes by o $\leftrightarrow$ p, r $\leftrightarrow$ s, and q $\leftrightarrow$ t, and the forbidden processes by p $\leftrightarrow$ q, and q $\leftrightarrow$ r. Accordingly, the non-extremal initial state $\mathbf{i}$, is connected with the final states, by the adiabatic solution curves
\begin{equation}\label{eq:piecewise}
  y(x) = \left\{
             \begin{array}{ll}
             y_a(x),& \quad \textrm{for}\,\, x_{0}\leq x<x_{i}, \\
             \medskip\\
             y_b(x),& \quad \textrm{for}\,  x_{i}\leq x<\infty. \\
             \end{array} \right.
\end{equation}
In this sense, we can ramify the physically allowed branches of the solutions, as shown in Fig. \ref{fig3}. In this diagram, the physically accepted parts of the solution $y(x)$, are those that connect, adiabatically, the initial state $\mathbf{i}\equiv (x_i, y_i)$, with $y_i<x_i$,  to another state $\mathbf{f}\equiv (x_f, y_f)$, with $y_f<x_f$, following the $y_a$ branch, if $x_f<x_i$, and the $y_b$ branch, if $x_f>x_i$. 
}

\begin{figure}[h!]
	\begin{center}
		\includegraphics[width=85mm]{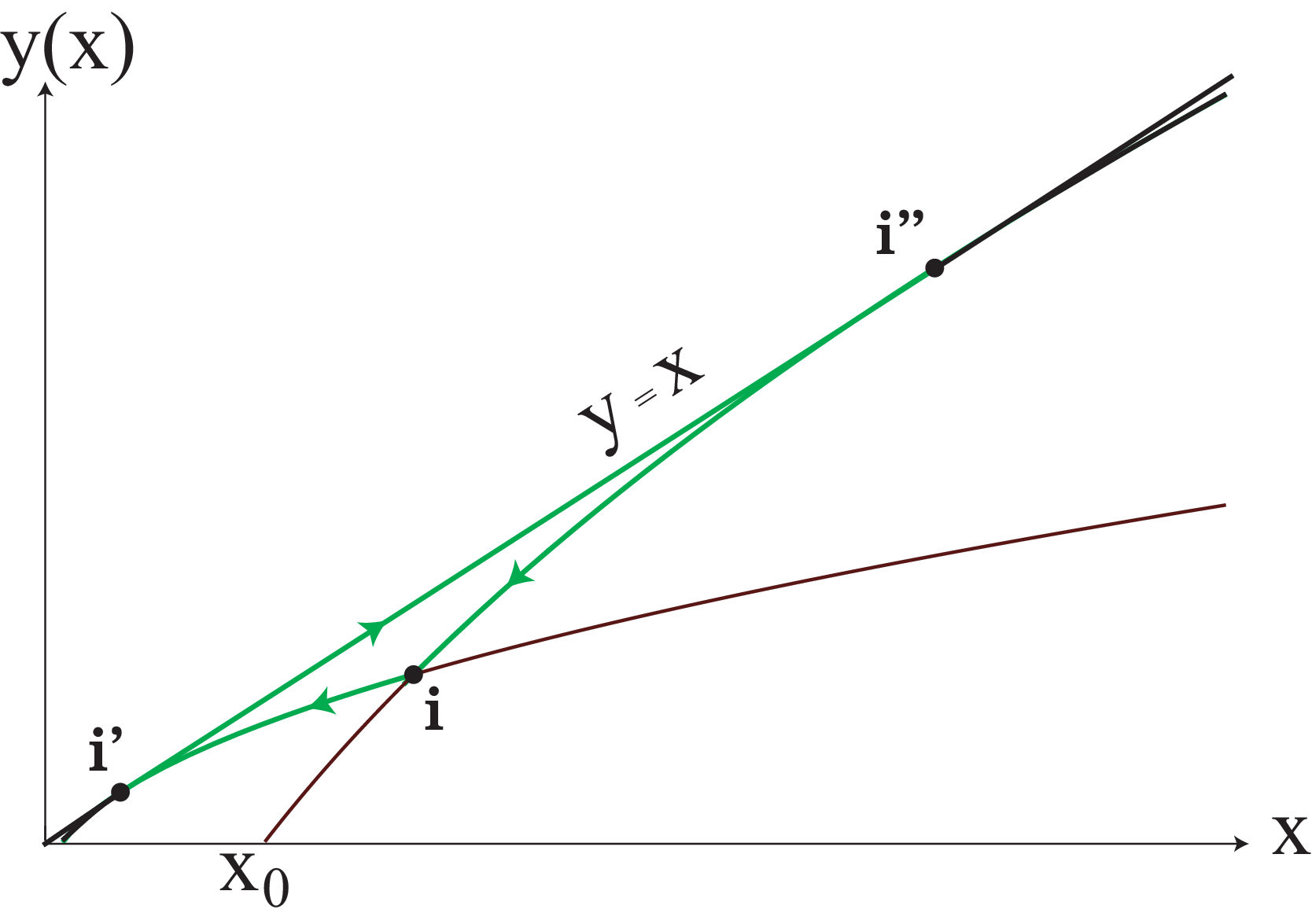}
	\end{center}
	\caption{The Carath\'{e}odory's postulate, prevents the formation of an adiabatic cycle. In such cycle, the equilibrium state $\mathbf{i}$ is adiabatically connected, respectively, to the extremal states $\mathbf{i}'$ and $\mathbf{i}''$, and then, is returned to itself (see the green triangular path). Such cycle is prohibited, because the states $\mathbf{i}'$ and $\mathbf{i}''$ are inaccessible for the state $\mathbf{i}$.}
	\label{fig4}
\end{figure}

A direct consequence of the condition 3, is that it prevents forming an adiabatic cycle (as desired by engineering). Such cycle has been illustrated in Fig. \ref{fig4}. Referring to the triangular path in this diagram, the state $\mathbf{i}$, residing in the submanifold $\mathcal{T}>0$, is first connected, adiabatically, to the state $\mathbf{i}'$, and then to the state $\mathbf{i}''$, which both reside in the submanifold $\mathcal{T}=0$, and finally, it is returned to $\mathbf{i}$. Note that, the process $\mathbf{i}'\rightarrow  \mathbf{i}''$ is adiabatic and isothermal. In fact, since these states are inaccessible by the Carath\'{e}odory's postulate, the above adiabatic cycle is not allowed to form.

\begin{figure}[h!]
    \centering
    \includegraphics[width=9.5cm]{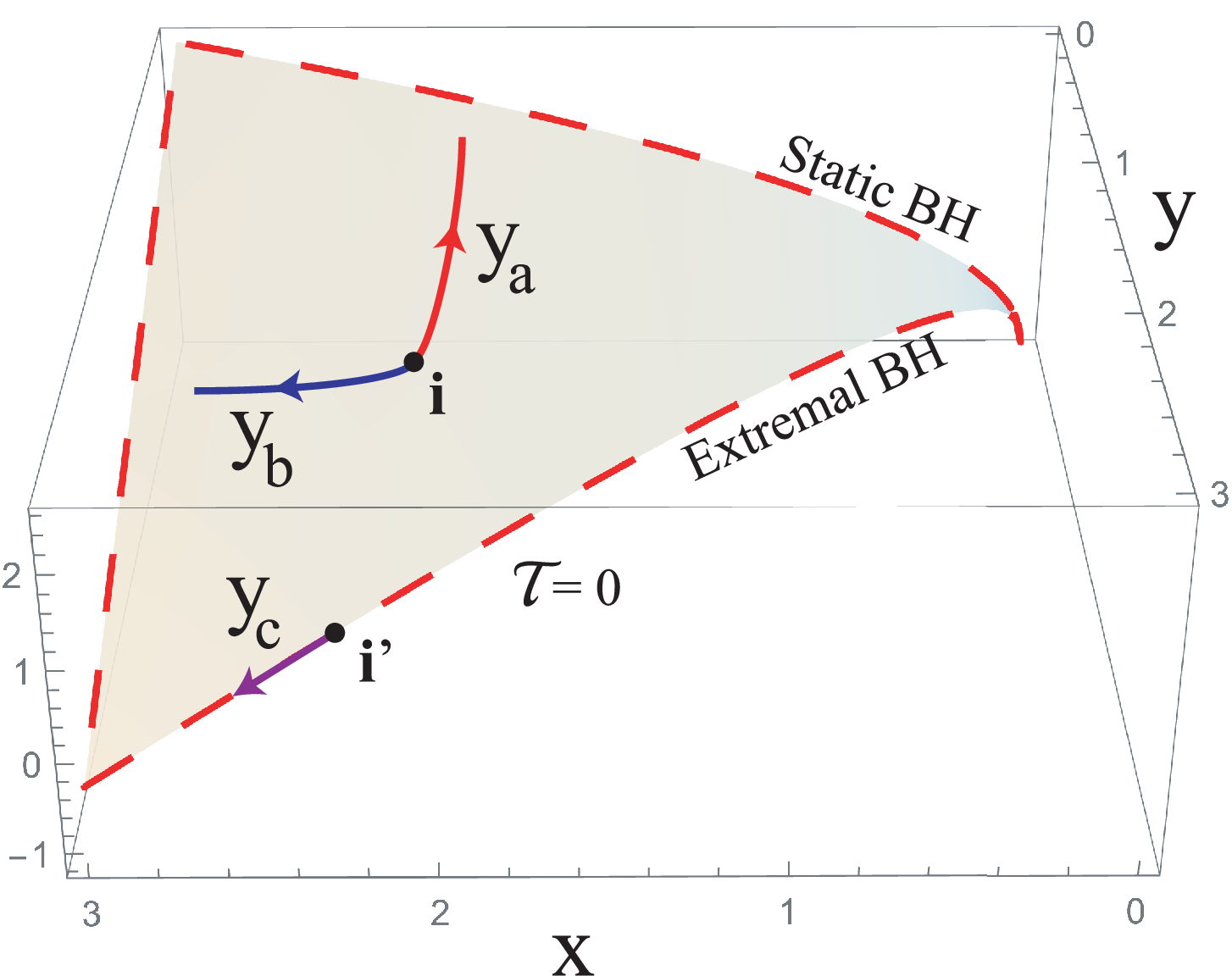}
    \caption{The adiabatic surface plotted for $\zeta_a = 1.2$, which is confined by the extremal, and the static black hole (BH) limits. The adiabatic processes $y_a$ and $y_b$, can connect the initial state $\mathbf{i}$ to other points on the surface. Same holds for the adiabatic process $y_c$ on the extremal limit, connecting the initial point $\mathbf{i}'$ to other points on the same line. The transmission $\mathbf{i}\rightarrow \mathbf{i}'$ is, however, prohibited by the Carath\'{e}odory's postulate.
    }
    \label{fig:M_ya_region}
\end{figure}

{Accordingly, the extremal limit should be excluded from the adiabatic hypersurface, since there is no adiabatic process that can reach this state. In fact, either of the solutions \eqref{sol1} and \eqref{sol2} can produce an adiabatic surface, that lies between the extremal $(y=x)$ and the static $(y=0)$ black hole limits (see Fig.~\ref{fig:M_ya_region}).
}
\textcolor{Blue}{}

Note that, since the conditional solution given by Eq. (\ref{eq:piecewise}) excludes the extremal states (the $\mathcal{T} = 0$ leaf), we can ensure the satisfaction of the third law through all isoperimetral (or adiabatic) processes.

\section{Scattering of two extremal black holes and the second law}\label{sec4}

{In this section, we explore the possibility of occurring an isolated merger of two extremal BTZ black holes with the initial states $(\mathcal{M}_1, \mathcal{J}_1)$ and $(\mathcal{M}_2, \mathcal{J}_2)$, to produce the final state $(\mathcal{M}_1+\mathcal{M}_2, \mathcal{J}_1+\mathcal{J}_2)$. As in Refs. \cite{Belgiorno_Martellini,Molina:2021hgx,Fathi:2021bxk}, we define the quantity
	\begin{equation}
	\alpha^{2} (\mathcal{M}, \mathcal{J})=\mathcal{M}^{2}- \mathcal{J}^{2},  \label{cs1}
	\end{equation}%
	so that, for the initial black holes we have $\alpha_{\mathrm{in}}=\alpha_{1}+\alpha_{2}$, where
\begin{subequations}\label{cs2}
\begin{align}
	 &  \alpha _{1}^{2}(\mathcal{M}_1, \mathcal{J}_1)=\mathcal{M}_1^{2}- \mathcal{J}_1^{2}\geq 0,\\
	 &  \alpha _{2}^{2}(\mathcal{M}_2, \mathcal{J}_2)=\mathcal{M}_2^{2}- \mathcal{J}_2^{2}\geq 0, 
\end{align}
\end{subequations}
and for the final black hole, $\alpha_{\mathrm{fin}}=\alpha_{12}$, with
\begin{multline}\label{cs3}
	\alpha _{12}^{2} (\mathcal{M}_1+\mathcal{M}_2, \mathcal{J}_1+\mathcal{J}_2)=\left(
	\mathcal{M}_1+\mathcal{M}_2\right)^{2}-\left(\mathcal{J}_1+\mathcal{J}_2\right)^{2} \\
	= \alpha _{1}^{2}(\mathcal{M}_1, \mathcal{J}_1) +	\alpha _{1}^{2}(\mathcal{M}_2, \mathcal{J}_2) +2\left( \mathcal{M}_1\,\mathcal{M}_2-\mathcal{J}_1\,\mathcal{J}_2\right).
\end{multline}}{Therefore, for two extremal black holes of the initial states $\alpha_{1}^{2}(\mathcal{M}_{\mathrm{\mathrm{ext}}}^{(1)}, \mathcal{J}_{\mathrm{ext}}^{(1)})=\alpha _{2}^{2}(\mathcal{M}_{\mathrm{ext}}^{(2)}, \mathcal{J}_{\mathrm{ext}}^{(2)})=0$, and of the positive masses $\mathcal{M}_{\mathrm{ext}}=|\mathcal{J}_{\mathrm{ext}}|>0$, Eq.~\eqref{cs3} becomes
\begin{eqnarray}\label{cs4}
	\alpha _{12}^{2}&&\left(\mathcal{M}_{\mathrm{ext}}^{(1)}+\mathcal{M}_{\mathrm{ext}}^{(2)}, \mathcal{J}_{\mathrm{ext}}^{(1)}+\mathcal{J}_{\mathrm{ext}}^{(2)}\right)\nonumber\\
	&=&2\left( \mathcal{M}_{\mathrm{ext}}^{(1)}\,\mathcal{M}_{\mathrm{ext}}^{(2)}-\mathcal{J}_{\mathrm{ext}}^{(1)}\,\mathcal{J}_{\mathrm{ext}}^{(2)}\right)\nonumber \\
	&=&2\left( \left|\mathcal{J}_{\mathrm{ext}}^{(1)}\right|\,\left|\mathcal{J}_{\mathrm{ext}}^{(2)}\right|-\mathcal{J}_{\mathrm{ext}}^{(1)}\,\mathcal{J}_{\mathrm{ext}}^{(2)}\right)\nonumber \\
&=&2\left|\mathcal{J}_{\mathrm{ext}}^{(1)}\right|\,\left|\mathcal{J}_{\mathrm{ext}}^{(2)}\right| \left( 1-\cos \beta \right), 
\end{eqnarray}%
where $\beta=\measuredangle(\mathcal{J}_{\mathrm{ext}}^{(1)}, \mathcal{J}_{\mathrm{ext}}^{(2)})$. Thus, for extremal black holes rotating in the same direction ($\beta=0$), the final black hole will be, as well, an extremal one. This is while, for those rotating in opposite directions  ($\beta=\pi$), the final black hole will not be extremal.
} 

{In the case of having an extremal final state, a possible violation of the second law may occur, since, in fact, the process is irreversible and the entropy of the final state should be greater than that of the initial states (under the assumption that the system is isolated and there is no exchange of energy with the rest of the universe). On the other hand, if the final state is non-extremal, then its entropy is, naturally, greater than that of the initial states. 
As it was formerly given in Eq.~\eqref{entrbtz}, the entropy of the BTZ black hole is $\mathcal{S}=g \mathcal{P}/4$, where $\mathcal{P}$ is the perimeter of the event horizon and $g$ is the \textit{genus}, that characterizes the topology of the thermodynamic manifold. In this sense, the extremal and non-extremal black holes are characterized, respectively, by $g=0$ and $g=1$. Hence, the latter corresponds to the change of the topology of the black hole.}

In fact, passing from one black hole topology to another, we encounter the spacetime singularities. And since the above processes accure in the classical environment, these singularities are inevitable. 
Therefore, to avoid the complexities associated with the change in the topology, it is convenient to infer that the scattering of two initially extremal BTZ black holes  leads to an extremal BTZ black hole, and this, violates the second law. The above statements can be summarized as
\begin{equation}\label{eq:beta}
  \beta = \left\{
             \begin{array}{ll}
            0~~\Rightarrow~~ \mathrm{violation~of~the~second~law}, \
             \medskip\\
             \pi~~\Rightarrow~~ \mathrm{change~of~the~black~hole~topology}. \\
             \end{array} \right.
\end{equation}

\section{Discussion and final remarks}\label{finrem}

{In this paper we studied the thermodynamics of the BTZ black hole, based the axiomatic approach of Carath\'{e}odory. We introduced the Pfaffian form $\delta Q_{{\rm rev}}$, to represent the infinitesimal heat exchanged reversibly, that gives a definition to the metric entropy and temperature, with the latter, as an integrating factor for the Pfaffian form.} 
{The natural extensive variables of the uncharged BTZ black hole in the equilibrium thermodynamics space  (i.e. the homogeneous variables of degree one), are $(\mathcal{M}, \mathcal{J})$, that has an associated Pfaffian form $\delta Q_{{\rm rev}} = \mathrm{d}\mathcal{M}-\mathcal{W} \mathrm{d} \mathcal{J}$. The symmetry of the homogeneity for $\delta Q_{{\rm rev}}$, can then be inspected by means of the Euler vector field (Liouville operator) in Eq.~\eqref{eulervec}, which indicates the consistency of the methods given here, with the thermodynamic definition of the temperature.
}

{As the first application of the presented approach, we studied the adiabatic processes, by analyzing the corresponding Cauchy problem.
In this sense, the problem is equivalent to the adiabatic processes in the Reissner-Nordstr\"om (RN) black hole spacetime \cite{Belgiorno_Martellini}, since, regarding the adiabatic transformations, the electric charge for the RN black hole plays the same role as the angular momentum for the BTZ black hole. We will address this issue in a future work.}

{Since the obtained adiabatic solutions allow for two definite constants, they can be therefore employed in the correct physical description of the  acceptable adiabatic paths. This way, and to respect the second, and especially the third law, the extremal submanifold ($\mathcal{T}=0$) must be disconnected from the non-extremal one ($\mathcal{T}>0$). In fact, the thermodynamic foliation of the non-extremal states, allows us to have a consistent construction of thermodynamics, since there are proven arguments to connect the approaches of Carath\'eodory and Gibbs \cite{Belgiorno2002Bridge}. Consequently, the Hawking-Bekenstein entropy formula is found valid, only for the non-extremal states. The entropy of the extremal states is, on the other hand, considered to be zero.
}

The classical merging of two extremal rotating BTZ black holes, provided us another tool to inspect the leaf $\mathcal{T}=0$,  and the corresponding  property $\mathcal{S}=0$. The unconformity of the second law with the aforementioned entropy condition, necessitates the inclusion of a net electrical charge for the black hole. In this sense, a new equivalence with the RN black hole could be found, which in that case, relaxes the problem by introducing a definite angular momentum to the system \cite{Belgiorno_Martellini}.
%

\begin{acknowledgements}
M. Fathi has been supported by the Agencia Nacional de Investigaci\'{o}n y Desarrollo (ANID) through DOCTORADO Grants No. 2019-21190382, and No. 2021-242210002. J.V. was partially supported by the Centro de Astrof\'isica de Valpara\'iso (CAV).	
\end{acknowledgements}

\bibliographystyle{ieeetr} 
\bibliography{Biblio_v1.bib}

\begin{thebibliography}{10}

\bibitem{Achucarro:1986}
A.~Achúcarro and P.~Townsend, ``{A {Chern}-{Simons} action for
  three-dimensional anti-de {Sitter} supergravity theories},'' {\em Phys. Lett.
  B}, vol.~180, no.~1-2, pp.~89--92, 1986.

\bibitem{Witten:1988}
{Witten, Edward}, ``{2+1 dimensional gravity as an exactly soluble system},''
  {\em {Nucl. Phys. B}}, vol.~311, no.~1, pp.~46--78, 1988.

\bibitem{Banados:1992wn}
M.~Ba\~nados, C.~Teitelboim, and J.~Zanelli, ``{The black hole in
  three-dimensional space-time},'' {\em Phys. Rev. Lett.}, vol.~69,
  pp.~1849--1851, 1992.

\bibitem{Banados:1992gq}
M.~Ba\~nados, M.~Henneaux, C.~Teitelboim, and J.~Zanelli, ``{Geometry of the
  (2+1) black hole},'' {\em Phys. Rev. D}, vol.~48, pp.~1506--1525, 1993.
\newblock [Erratum: Phys. Rev.D88,069902(2013)].

\bibitem{Carlip:1995qv}
S.~Carlip, ``{The (2+1)-Dimensional black hole},'' {\em Class. Quant. Grav.},
  vol.~12, pp.~2853--2880, 1995.

\bibitem{Banados:1998gg}
M.~Ba\~nados, ``{Three-dimensional quantum geometry and black holes},'' {\em
  AIP Conf. Proc.}, vol.~484, no.~1, pp.~147--169, 1999.

\bibitem{Cataldo:2000qi}
M.~Cataldo, S.~del Campo, and A.~A. Garc\'ia, ``{BTZ black hole from (3+1)
  gravity},'' {\em Gen. Rel. Grav.}, vol.~33, pp.~1245--1255, 2001.

\bibitem{AyonBeato:2004if}
E.~Ayon-Beato, C.~Martinez, and J.~Zanelli, ``{Birkhoff's theorem for
  three-dimensional AdS gravity},'' {\em Phys. Rev. D}, vol.~70, p.~044027,
  2004.

\bibitem{Witten:2007}
E.~{Witten}, ``{Three-dimensional gravity revisited},'' {\em arXiv e-prints},
  p.~arXiv:0706.3359, 2007.

\bibitem{Cruz_1994}
N.~Cruz, C.~Mart{\'{\i}}nez, and L.~Pe{\~{n}}a, ``Geodesic structure of the
  (2+1)-dimensional {BTZ} black hole,'' {\em Class. Quant. Grav.}, vol.~11,
  no.~11, pp.~2731--2739, 1994.

\bibitem{Gamboa:2000uc}
J.~Gamboa and F.~M\'endez, ``{Scattering in three dimensional extremal black
  holes},'' {\em Class. Quant. Grav.}, vol.~18, pp.~225--232, 2001.

\bibitem{Lepe:2003na}
S.~Lepe, F.~M\'endez, J.~Saavedra, and L.~Vergara, ``{Fermions scattering in a
  three-dimensional extreme black hole background},'' {\em Class. Quant.
  Grav.}, vol.~20, pp.~2417--2428, 2003.

\bibitem{Cardoso:2001hn}
V.~Cardoso and J.~P.~S. Lemos, ``{Scalar, electromagnetic and Weyl
  perturbations of BTZ black holes: Quasinormal modes},'' {\em {Phys. Rev. D}},
  vol.~63, p.~124015, 2001.

\bibitem{Birmingham:2003wa}
D.~Birmingham, S.~Carlip, and Y.-J. Chen, ``{Quasinormal modes and black hole
  quantum mechanics in (2+1)-dimensions},'' {\em Class. Quant. Grav.}, vol.~20,
  pp.~L239--L244, 2003.

\bibitem{Crisostomo:2004hj}
J.~Crisostomo, S.~Lepe, and J.~Saavedra, ``{Quasinormal modes of extremal BTZ
  black hole},'' {\em Class. Quant. Grav.}, vol.~21, pp.~2801--2810, 2004.

\bibitem{Setare:2003hm}
M.~R. Setare, ``{Nonrotating BTZ black hole area spectrum from quasinormal
  modes},'' {\em Class. Quant. Grav.}, vol.~21, pp.~1453--1458, 2004.

\bibitem{Cruz:1994ar}
N.~Cruz and J.~Zanelli, ``{Stellar equilibrium in (2+1)-dimensions},'' {\em
  Class. Quant. Grav.}, vol.~12, pp.~975--982, 1995.

\bibitem{Garcia:2004jz}
A.~A. Garc\'ia, ``{Stationary circularly symmetric 2+1 rigidly rotating perfect
  fluids},'' {\em Phys. Rev. D}, vol.~69, p.~124024, 2004.

\bibitem{Garcia_cuath}
A.~A. Garc\'ia and C.~Campuzano, ``{All static circularly symmetric perfect
  fluid solutions of (2+1) gravity},'' {\em Phys. Rev. D}, vol.~67, p.~064014,
  2003.

\bibitem{COV_2004BTZ}
N.~Cruz, M.~Olivares, and J.~R. Villanueva, ``{Static circularly symmetric
  perfect fluid solutions with an exterior BTZ metric},'' {\em Gen. Rel.
  Grav.}, vol.~37, pp.~667--674, 2005.

\bibitem{Gundlach:2020ovt}
C.~Gundlach and P.~Bourg, ``{Rigidly rotating perfect fluid stars in $2+1$
  dimensions},'' {\em Phys. Rev. D}, vol.~102, no.~8, p.~084023, 2020.

\bibitem{Rincon:2018I}
A.~Rinc\'on and B.~Koch, ``{Scale-dependent BTZ black hole},'' {\em Eur. Phys.
  J. C}, vol.~78, no.~12, p.~1022, 2018.

\bibitem{Rincon:2018II}
A.~Rinc\'on, E.~Contreras, P.~Bargue\~no, B.~Koch, and G.~Panotopoulos,
  ``{Scale-dependent ( $2+1$ )-dimensional electrically charged black holes in
  Einstein-power-Maxwell theory},'' {\em Eur. Phys. J. C}, vol.~78, no.~8,
  p.~641, 2018.

\bibitem{Rincon:2019zxk}
A.~Rinc\'on and J.~R. Villanueva, ``{The Sagnac effect on a scale-dependent
  rotating BTZ black hole background},'' {\em Class. Quant. Grav.}, vol.~37,
  no.~17, p.~175003, 2020.

\bibitem{Fathi:2019jid}
M.~Fathi, A.~Rinc\'on, and J.~R. Villanueva, ``{Photon trajectories on a first
  order scale-dependent static BTZ black hole},'' {\em Class. Quant. Grav.},
  vol.~37, no.~7, p.~075004, 2020.

\bibitem{Rincon:2020izv}
A.~Rinc\'on, E.~Contreras, F.~Tello-Ort\'iz, P.~Bargue\~no, and G.~Abell\'an,
  ``{Anisotropic 2+1 dimensional black holes by gravitational decoupling},''
  {\em Eur. Phys. J. C}, vol.~80, no.~6, p.~490, 2020.

\bibitem{Vagenas:2004zt}
E.~C. Vagenas, ``{Energy distribution in a BTZ black hole spacetime},'' {\em
  Int. J. Mod. Phys. D}, vol.~14, pp.~573--586, 2005.

\bibitem{Cataldo_Garcia}
M.~Cataldo and A.~A. Garc\'ia, ``{Regular (2+1)-dimensional black holes within
  nonlinear electrodynamics},'' {\em Phys. Rev. D}, vol.~61, p.~084003, 2000.

\bibitem{Cataldo_Salgado}
M.~Cataldo and P.~Salgado, ``{Three dimensional extreme black hole with self
  (anti-self) dual Maxwell field},'' {\em Phys. Lett. B}, vol.~448, pp.~20--25,
  1999.

\bibitem{Carlip:2005zn}
S.~Carlip, ``{Conformal field theory, (2+1)-dimensional gravity, and the BTZ
  black hole},'' {\em Class. Quant. Grav.}, vol.~22, pp.~R85--R124, 2005.

\bibitem{Cai:1996df}
R.~G. Cai, Z.~J. Lu, and Y.~Z. Zhang, ``{Critical behavior in (2+1)-dimensional
  black holes},'' {\em Phys. Rev. D}, vol.~55, pp.~853--860, 1997.

\bibitem{Banados:1998ta}
M.~Ba\~nados, T.~Brotz, and M.~E. Ort\'iz, ``{Boundary dynamics and the
  statistical mechanics of the (2+1)-dimensional black hole},'' {\em Nucl.
  Phys. B}, vol.~545, pp.~340--370, 1999.

\bibitem{Wang:2006eb}
S.~Wang, S.-Q. Wu, F.~Xie, and L.~Dan, ``{The First laws of thermodynamics of
  the (2+1)-dimensional BTZ black holes and Kerr-de Sitter spacetimes},'' {\em
  Chin. Phys. Lett.}, vol.~23, pp.~1096--1098, 2006.

\bibitem{Dolan:2010ha}
B.~P. Dolan, ``{The cosmological constant and the black hole equation of
  state},'' {\em Class. Quant. Grav.}, vol.~28, p.~125020, 2011.

\bibitem{Sarkar_2006}
T.~Sarkar, G.~Sengupta, and B.~N. Tiwari, ``On the thermodynamic geometry of
  {BTZ} black holes,'' {\em {J. High Energy Phys.}}, vol.~2006, no.~11,
  pp.~015--015, 2006.

\bibitem{Quevedo:2008ry}
H.~Quevedo and A.~Sanchez, ``{Geometric description of BTZ black holes
  thermodynamics},'' {\em Phys. Rev. D}, vol.~79, p.~024012, 2009.

\bibitem{Akbar:2011qw}
M.~Akbar, H.~Quevedo, K.~Saifullah, A.~Sanchez, and S.~Taj, ``{Thermodynamic
  geometry of charged rotating BTZ lack holes},'' {\em Phys. Rev. D}, vol.~83,
  p.~084031, 2011.

\bibitem{Hendi2}
S.~H. Hendi, S.~Panahiyan, B.~E. Panah, and M.~Momennia, ``A new approach
  toward geometrical concept of black hole thermodynamics,'' {\em The European
  Physical Journal C}, vol.~75, p.~507, Oct. 2015.

\bibitem{Hendi1}
S.~H. Hendi, B.~E. Panah, and S.~Panahiyan, ``Massive charged {BTZ} black holes
  in asymptotically (a){dS} spacetimes,'' {\em Journal of High Energy Physics},
  vol.~2016, p.~29, May 2016.

\bibitem{Singh_2014}
D.~V. Singh and S.~Siwach, ``Thermodynamics of {BTZ} black hole and
  entanglement entropy,'' {\em J. Phys.: Conference Series}, vol.~481,
  p.~012014, 2014.

\bibitem{Alsaleh}
S.~{Alsaleh}, ``{Thermodynamics of BTZ black holes in gravity{\textquoteright}s
  rainbow},'' {\em Int. J. Mod. Phys. A}, vol.~32, no.~15, p.~1750076, 2017.

\bibitem{DEHGHANI2018351}
M.~Dehghani, ``Thermodynamics of charged dilatonic btz black holes in rainbow
  gravity,'' {\em {Phys. Lett. B}}, vol.~777, pp.~351--360, 2018.

\bibitem{Liang:2019jnj}
T.~Liang, W.~Tang, and W.~Xu, ``{Entropy relations and bounds of BTZ black hole
  in gravity's rainbow},'' {\em Int. J. Mod. Phys. D}, vol.~28, no.~08,
  p.~1950109, 2019.

\bibitem{Camci:2020yre}
U.~Camci, ``{Three-dimensional black holes via Noether symmetries},'' {\em
  Phys. Rev. D}, vol.~103, no.~2, p.~024001, 2021.

\bibitem{Chougule:2018cny}
S.~Chougule, S.~Dey, B.~Pourhassan, and M.~Faizal, ``{BTZ black holes in
  massive gravity},'' {\em Eur. Phys. J. C}, vol.~78, no.~8, p.~685, 2018.

\bibitem{Bravo-Gaete:2014haa}
M.~Bravo-Gaete and M.~Hassaine, ``{Thermodynamics of a BTZ black hole solution
  with an Horndeski source},'' {\em Phys. Rev. D}, vol.~90, no.~2, p.~024008,
  2014.

\bibitem{Ortiz:2018ddt}
L.~Ort\'iz and N.~Bret\'on, ``{Aspects of the BTZ black hole interacting with
  fields},'' {\em Mod. Phys. Lett. A}, vol.~34, no.~31, p.~1950251, 2019.

\bibitem{Townsend:2013ela}
P.~K. Townsend and B.~Zhang, ``{Thermodynamics of ''Exotic''
  Ba\~nados-Teitelboim-Zanelli Black Holes},'' {\em Phys. Rev. Lett.},
  vol.~110, no.~24, p.~241302, 2013.

\bibitem{Kiselev:2005jb}
V.~V. Kiselev, ``{Entropy of BTZ black hole and its spectrum by quantum radial
  geodesics behind horizons},'' {\em Phys. Rev. D}, vol.~73, p.~104018, 2006.

\bibitem{caratheodory09}
C.~Carath\'eodory, ``{Untersuchungen \"uber die Grundlagen der
  Thermodynamik.},'' {\em Math. Ann.}, vol.~67, pp.~355--386, 1909.

\bibitem{Belgiorno:2002BHCth}
F.~Belgiorno, ``{Black hole thermodynamics in Caratheodory's approach},'' {\em
  Phys. Lett. A}, vol.~312, pp.~324--330, 2003.

\bibitem{Belgiorno_Cacciatori}
F.~Belgiorno and S.~L. Cacciatori, ``{General symmetries: From homogeneous
  thermodynamics to black holes},'' {\em Eur. Phys. J. Plus}, vol.~126, p.~86,
  2011.

\bibitem{Belgiorno_Martellini}
F.~Belgiorno and M.~Martellini, ``Black holes and the third law of
  thermodynamics,'' {\em Int. J. Mod. Phys. D}, vol.~13, pp.~739--770, 2004.

\bibitem{Cruzlepe04}
N.~Cruz and S.~Lepe, ``{On the thermal description of the BTZ black holes},''
  {\em Phys. Lett. B}, vol.~593, pp.~235--241, 2004.

\bibitem{buchdahl49}
H.~A. Buchdahl, ``On the principle of carathéodory,'' {\em Am. J. Phys.},
  vol.~17, no.~1, pp.~41--43, 1949.

\bibitem{buchdahl49I}
H.~A. Buchdahl, ``On the theorem of carathéodory,'' {\em Am. J. Phys.},
  vol.~17, no.~1, pp.~44--46, 1949.

\bibitem{buchdahl49II}
H.~A. Buchdahl, ``On the unrestricted theorem of carathéodory and its
  application in the treatment of the second law of thermodynamics,'' {\em Am.
  J. Phys.}, vol.~17, no.~4, pp.~212--218, 1949.

\bibitem{buchdahl54}
H.~A. Buchdahl, ``Integrability conditions and carathéodory's theorem,'' {\em
  Am. J. Phys.}, vol.~22, no.~4, pp.~182--183, 1954.

\bibitem{buchdahl55}
H.~A. Buchdahl, ``Simplification of a proof of carathéodory's theorem,'' {\em
  Am. J. Phys.}, vol.~23, no.~1, pp.~65--66, 1955.

\bibitem{Landsberg1964ADO}
P.~T. Landsberg, ``A deduction of carath{\'e}odory's principle from kelvin's
  principle,'' {\em Nature}, vol.~201, pp.~485--486, 1964.

\bibitem{Marshall78}
T.~W. {Marshall}, ``{A simplified version of Carath{\'e}odory
  thermodynamics},'' {\em Am. J. Phys.}, vol.~46, no.~2, pp.~136--137, 1978.

\bibitem{boyling68}
J.~Boyling, ``{Carathéodory's principle and the existence of global
  integrating factors},'' {\em Commun. Math. Phys.}, vol.~10, p.~52–68, 1968.

\bibitem{boyling72}
J.~Boyling, ``{An axiomatic approach to classical thermodynamics},'' {\em Proc.
  R. Soc. Lond. A.}, vol.~329, p.~35–70, 1972.

\bibitem{pogliani}
L.~Pogliani and M.~Berberan-Santos, ``{Constantin Carathéodory and the
  axiomatic thermodynamics},'' {\em J. Math. Chem.}, vol.~28, p.~313–324,
  2000.

\bibitem{Belgiorno2002Bridge}
F.~{Belgiorno}, ``{Homogeneity as a bridge between Carath{\'e}odory and
  Gibbs},'' {\em arXiv e-prints}, pp.~math--ph/0210011, 2002.

\bibitem{Belgiorno:2002QTBH}
F.~Belgiorno, ``{Quasihomogeneous thermodynamics and black holes},'' {\em J.
  Math. Phys.}, vol.~44, pp.~1089--1128, 2003.

\bibitem{Belgiorno_2003aNTL}
F.~{Belgiorno}, ``{Notes on the third law of thermodynamics: I},'' {\em J.
  Phys. A: Math. Gen.}, vol.~36, no.~30, pp.~8165--8193, 2003.

\bibitem{Belgiorno_2003bNTL}
F.~Belgiorno, ``{Notes on the third law of thermodynamics: II},'' {\em J. Phys.
  A: Math. Gen.}, vol.~36, no.~30, pp.~8195--8221, 2003.

\bibitem{Hawking:1994ii}
S.~W. Hawking, G.~T. Horowitz, and S.~F. Ross, ``{Entropy, Area, and black hole
  pairs},'' {\em Phys. Rev. D}, vol.~51, pp.~4302--4314, 1995.

\bibitem{Teitelboim:1994az}
C.~Teitelboim, ``{Action and entropy of extreme and nonextreme black holes},''
  {\em Phys. Rev. D}, vol.~51, p.~4315, 1995.
\newblock [Erratum: Phys.Rev.D 52, 6201 (1995)].

\bibitem{carroll_extremal_2009}
S.~M. Carroll, M.~C. Johnson, and L.~Randall, ``Extremal limits and black hole
  entropy,'' {\em J. High Energy Phys.}, vol.~2009, no.~11, pp.~109--109, 2009.

\bibitem{lemos_entropy_2016}
J.~P. Lemos, G.~M. Quinta, and O.~B. Zaslavskii, ``Entropy of extremal black
  holes: {Horizon} limits through charged thin shells in a unified approach,''
  {\em Phys. Rev. D}, vol.~93, no.~8, p.~084008, 2016.

\bibitem{lemos_entropy_2014}
J.~P. Lemos and G.~M. Quinta, ``Entropy of thin shells in a (2+1)-dimensional
  asymptotically {AdS} spacetime and the {BTZ} black hole limit,'' {\em Phys.
  Rev. D}, vol.~89, no.~8, p.~084051, 2014.

\bibitem{lemos_thermodynamics_2015}
J.~P. Lemos, F.~J. Lopes, M.~Minamitsuji, and J.~V. Rocha, ``{Thermodynamics of
  rotating thin shells in the {BTZ} spacetime},'' {\em Phys. Rev. D}, vol.~92,
  no.~6, p.~064012, 2015.

\bibitem{lemos_unified_2017}
J.~P. Lemos, M.~Minamitsuji, and O.~B. Zaslavskii, ``Unified approach to the
  entropy of an extremal rotating {BTZ} black hole: {Thin} shells and horizon
  limits,'' {\em Phys. Rev. D}, vol.~96, no.~8, p.~084068, 2017.

\bibitem{Molina:2021hgx}
M.~Molina and J.~R. Villanueva, ``{On the thermodynamics of the Hayward black
  hole},'' {\em Class. Quant. Grav.}, vol.~38, no.~10, p.~105002, 2021.

\bibitem{Fathi:2021bxk}
M.~Fathi, M.~Molina, and J.~R. Villanueva, ``Adiabatic evolution of hayward
  black hole,'' {\em Phys. Lett. B}, vol.~820, no.~136548, 2021.

\end{thebibliography}

\end{document}